\documentclass[twocolumn,prl,showpacs,amsmath,amssymb]{revtex4}
\usepackage[OT1]{fontenc}
\usepackage{amsfonts}
\usepackage{hyperref}
\def\be{\begin{equation}}
\def\ee{\end{equation}}
\def\bea{\begin{eqnarray}}
\def\eea{\end{eqnarray}}
\def\bma{\begin{mathletters}}
\def\ema{\end{mathletters}}

\def\0{\overline{0}}

\def\q0{\underline{0}}

\def\tr{\mbox{tr}}
\newcommand\one[1][]{\mathbb I}

\def\bra#1{\langle#1|} \def\ket#1{|#1\rangle}

\def\proj#1{\ket{#1}\!\bra{#1}}

\newcommand{\qph}[1]%
    {\href{http://arxiv.org/abs/quant-ph/#1}%
    {\texttt{quant-ph/#1}}}
\begin{document}

\title{Optimality of Gaussian Attacks in Continuous Variable Quantum Cryptography
}
\author{Miguel Navascu\'es$^1$, Fr\'ed\'eric Grosshans$^2$, and Antonio Ac\'\i n$^1$}

\affiliation{ $^1$ICFO-Institut de Ciencies Fotoniques,
Mediterranean Technology Park, 08860 Castelldefels
(Barcelona),Spain\\
$^2$Laboratoire de Photonique Quantique et Mol\'eculaire, UMR CNRS
8537, ENS Cachan, 61 avenue du Pr\'esident Wilson, 94235 Cachan
Cedex, France}
\date{\today}


\begin{abstract}
We analyze the asymptotic security of the family of Gaussian
modulated Quantum Key Distribution protocols for Continuous
Variables systems. We prove that the Gaussian unitary attack is
optimal for all the considered bounds on the key rate when the
first and second momenta of the canonical variables involved are
known by the honest parties.
\end{abstract}


\maketitle



In 1984 Bennet and Brassard introduced the concept of Quantum
Cryptography and presented the first Quantum Key Distribution
(QKD) protocol: BB84 \cite{Bennet}. The original idea was that in
Quantum Mechanics, and contrary to Classical Physics, the
observation of a system invariably perturbs the system under
observation. Therefore, if two honest parties, Alice and Bob,
establish a quantum channel and use it to send information, an
eavesdropper's presence could be detected by analyzing how the
noise-free channel has changed.
It was then shown that QKD protocols are completely secure against any eavesdropping attacks as long as the bit
error rates do not exceed a certain value (see for instance \cite{Shor} and references therein). In the
meantime, new applications of Quantum Mechanics to certain information tasks started to develop: coin tossing,
dense coding, teleportation...

All these results first appeared in the context of discrete
systems, but many of them were later translated into the language
of Continuous Variables (CV) systems
. This is
{\sl per se} an interesting theoretical problem. However, the main
motivation for dealing with these systems comes from a practical
point of view:
although the set of feasible operations is reduced, the so-called
Gaussian operations are easy to implement and amazingly precise.
Quantum cryptography with continuous variables systems
\cite{GP,squeeze,CLvA,coher,coher2,het} was the most immediate
result: the transmission of coherent or squeezed pulses of light,
together with homodyne measurements, allows performing QKD with
very high key rates \cite{exper}.

The security analysis of these new protocols is not
straightforward. First of all, the commonly used reconciliation
and privacy amplification protocols are designed to correct and
distill secret bits from binary random variables, although some
have been adapted to continuous variables \cite{sliced, turbo}.
Second, the dimension of the Hilbert space on which the CV systems
are defined is infinite in theory, which makes a complete
tomography impossible in principle, thus preventing Alice and Bob
to know precisely the state they are actually sharing.
 Therefore,
security proofs for CV protocols have to consider the optimal
attack by Eve when Alice and Bob know their state is in some set,
usually defined by the momenta of the quadratures up to second
order \cite{gauss}. In her search for information, Eve's possible
attacks can be classified in three different types \cite{Kraus}:
\emph{individual attacks}, where Eve interacts individually with
the sent states and measures them individually before public
reconciliation; \emph{collective attacks}, where Eve applies the
same unitary individual attack over the sent states, but performs
her (possibly collective) measurements at any time during Alice
and Bob's reconciliation protocol and \emph{coherent attacks},
where Eve is allowed to perform any unitary collective interaction
over the sent states and any measurement strategy at any time she
wants. The latter is the most general attack Eve can use. Most of
the present security proofs give necessary and sufficient
conditions for key distillation when Eve is restricted to perform
an individual \cite{squeeze,CLvA} or finite-size coherent attack
\cite{gauss}. General proofs of security are given in \cite{GP}
for a squeezed-state protocol and in \cite{Assche,vAIC} for
coherent states.

Recently, bounds on extractable key rates have been derived for
the case of collective \cite{Winter,Renner} and general attacks
\cite{PhDRenner}. These bounds are easy to adapt to a wide class
of protocols, since they correspond to the difference of smooth
entropies, which tend to Von Neuman or Shannon entropies in the
asymptotic case. In this work we analyze a family of CV protocols
based on Gaussian modulation. This family includes most of the
protocols in the field of CV systems, such as those of Refs.
\cite{CLvA} using squeezed light, or those of Refs. \cite{coher,
het} that employ coherent states. We prove
that for all of them, the Gaussian attack is the unitary attack by Eve that minimizes the bounds on the key rate
of \cite{Winter,Renner}, when Alice and Bob know the quadrature momenta of their state up to the second order.
Therefore, Gaussian attacks turns out to be optimal for these protocols.

We consider quantum systems of $n$ canonical degrees of freedom,
called modes, belonging to $B({\cal H}({\mathbb R}^n))$. These are
characterized by the set of operators $\vec\Xi=(\Xi_1, \cdots,
\Xi_{2n})=(Q_1,P_1, \cdots, Q_n, P_n)$ satisfying the canonical
commutation relations $[\Xi_j,\Xi_k]=i(\sigma_n)_{jk}$, where
$\sigma_n$ is is the $n$-mode \emph{symplectic matrix},
\begin{equation}
\sigma_n=\bigoplus_{i=1}^n
\left(\begin{array}{cc}0&1\\-1&0\end{array}\right) .
\end{equation}
A state is said to be \emph{Gaussian} iff its density matrix, $\tilde\rho$, is the exponential of a quadratic
function $f$ on the canonical operators of the system, i.e.,
\begin{equation}
\label{rhoexpquad} \tilde\rho = \exp{[-f(\vec{\Xi})]}.
\end{equation}
Because of their simple structure, any Gaussian state can be
completely described in terms of its \emph{displacement vector},
$d$, and its \emph{covariance matrix}, $\gamma$, both defined as
\begin{equation}
    \begin{aligned}
    d_k&=\langle \Xi_k \rangle =\tr(\rho \,\Xi_k)\\
    \gamma_{kl}&=\tr\left(\rho \{\Xi_k-d_k,\Xi_l-d_l\}_{+}\right),
    \end{aligned}
\end{equation}
where $\{\}_{+}$ denotes the anti-commutator. Therefore, Gaussian
states are characterized just by the first and second order
momenta of the canonical variables $\vec\Xi$. \emph{Gaussian
operations} are completely positive (CP) maps that map Gaussian
states into Gaussian states.

The considered CV QKD protocols are based on random Gaussian
modulation of squeezed or coherent states of light
\cite{CLvA,coher,het}. They are \emph{prepare and measure}
(\emph{P\&M}) protocols, suitable to realistic implementation with
today's technology. However, for any \emph{P\&M} measure protocol
there exists a completely equivalent \emph{entanglement-based
scheme} \cite{BBM}. This description simplifies the theoretical
analysis, even if it would be more difficult to implement
experimentally. The entanglement-based scheme consists of the
following five steps (see also \cite{crypt2}): 1) Alice prepares a
two-mode squeezed state. 2) She performs a measurement over the
first mode. This measurement projects the second mode into a
randomly displaced (possibly squeezed) state. If Alice performs a
heterodyne measurement, she effectively prepares a coherent state
on the second mode. If she randomly chooses to perform a homodyne
measurement on $Q$ or $P$, she is effectively preparing a randomly
displaced squeezed state. 3) She sends the second mode to Bob via
a noisy quantum channel. 4) Bob receives the state sent by Alice.
He performs either a homodyne measurement in $Q$ or $P$, or a
heterodyne measurement, his result being $y$. 5) Alice and Bob
apply one-way Error Correction (EC) and Privacy Amplification (PA)
codes to distill a perfect secret key. If the classical
communication flows from Alice to Bob, we speak about \emph{Direct
Reconciliation (DR)}. On the contrary, if it is Bob who sends the
classical information to Alice during the reconciliation process,
we say they are using a \emph{Reverse Reconciliation (RR)}
protocol \cite{RR}.

Recently, general bounds on the extractable key rate under
collective attacks have been published \cite{Renner,Winter}. All
of them exploit the entanglement-based picture, but of course they
also apply to the corresponding \emph{P\&M} scheme. They are
expressed in terms of entropy quantities. Throughout this work,
the same notation $H$ is used for the (classical) Shannon entropy
and the (quantum) Von Neuman entropy. Let $X$ ($Y$) be the random
variable associated to Alice's (Bob's) measured quantity and by
$x$ ($y$) its value. According to \cite{Renner,Winter}, the key
rate $K$ obtained using Direct Reconciliation is bounded by
\begin{equation}
K\geq I(X:Y)-\chi(X:E)\equiv K_{\mbox{\scriptsize{coll}}} .
\label{Renner}
\end{equation}
Here $I(X:Y)$ denotes the classical mutual information,
$I(X:Y)=H(Y) - H(Y|X)$, while $\chi$ refers to the Holevo bound
\cite{Holevo},
\begin{gather}
    \label{Holevo}
    \chi(X:B)=H(B)-H(B|X) ,
\end{gather}
where $H(B|X)=\sum_x p(x) H(B|X=x)$. Formally, $I$ and $\chi$ look
identical, but they refer to different type of variables. While
the mutual information only deals with classical random variables,
the Holevo bound quantifies the accessible classical information
on quantum states. This justifies the different notation.

Suppose now that Bob is allowed to use a collective arbitrary measurement on many copies of the received states.
Of course, this is a rather unrealistic scenario, but it provides an upper bound to the maximum one-way secret
key rate when Bob is free to perform any \emph{individual} measurement. 
If, again, Eve is restricted to apply collective attacks, the key
rate, upon Bob optimizing his measurement, is given by
\cite{Winter}:
\begin{equation}
K\geq \chi(X:B)-\chi(X:E)\equiv K'_{\mbox{\scriptsize{coll}}} .
\label{Devetak}
\end{equation}
In these two bounds, namely Eqs. (\ref{Renner}) and
(\ref{Devetak}), the first term specifies the correlation between
the honest parties. It quantifies the amount of classical
information Alice and Bob should exchange to correct their errors.
The second term estimates Eve's knowledge on Alice (or Bob's)
variable. It is thus related to the amount of privacy
amplification required to make Eve's information vanishing.

Eve's attack has to be defined in order to compute the secret key
rate and needs therefore to be optimized. Indeed, after the
estimation strategy, Alice and Bob have some knowledge about their
state, this information being denoted by $g$. In the calculation
of key rates, as for any other interesting function, Alice and Bob
should minimize (\ref{Renner}) or \eqref{Devetak} over the set
$G$, consisting of all states $\rho$ compatible with $g$ (see also
\cite{WGC}).

In the CV scenario, it is natural to take $g$, i.e., Alice and
Bob's information on their state, as the first and second moments
on the measured quadratures. The first order correlations do not
play any role in the discussion, as they can be changed
arbitrarily by the use of local unitaries. As shown in the next
lines, for fixed second (and first) moments, the corresponding
Gaussian state optimizes the bounds on the key rates given above.
Interestingly, the Gaussian attack turns out to maximize     Eve's
information as well, $\chi(X:E)$. Before proceeding with the proof
of these results, we spend some lines clarifying the notation used
from now and on.

Let $\rho \in B({\cal H}^2)$ be a density matrix in any Hilbert
space $\cal H$. Then $\tilde{\rho}$ denotes the corresponding
density matrix of a Gaussian state characterized by the same
covariance matrix and displacement vector as $\rho$. Analogously,
if $p\,(\vec{x})$ is a probability distribution, then
$\tilde{p}\,(\vec{x})$ (or $\tilde p$ for short) denotes the
Gaussian probability distribution with the same first and second
momenta as $p\,(\vec{x})$. Moreover, if $F(\vec{x})$ represents
any quantity concerning a variable $\vec{x}$, described by a
certain distribution $p\,(\vec{x})$, then $\tilde{F}$ has to be
understood as the same functional $F$ calculated from the
distribution $\tilde{p}$. $\tilde\Delta F$ will be a shorthand
notation for the difference of these two quantities, $\tilde\Delta
F=\tilde F-F$.

Three results are used in what follows. First, let $\rho\in
B({\cal H}^2)$ be any physical state of a system $A$ and
$\bar\rho$ the one into which $\rho$ is transformed after the
measurement of the classical variable $X$. The measurement is
defined by a set of positive operators $\{M_x\}_x$ obeying $\sum_x
dx M_xM_x^\dagger=\one$. One has
\begin{equation}
\label{stmeas} \bar\rho=\sum_x \proj{x}dxM_x\rho M_x^\dagger=
\sum_x p\,(x)dx\proj{x}dx\otimes\rho_{|x},
\end{equation}
where $\rho|x$ is the normalized state of $\rho$ knowing $X=x$,
\begin{equation}
    \rho_{|x}=\frac{M_x\rho M_x^\dagger}{p\,(x)},
\end{equation}
and $p\,(x)=\tr(M_x\rho M_x^\dagger)$. It is straightforward to
check that
\begin{equation}
\label{descomp}
H(A|X)=H(\bar A)-H(X),
\end{equation}
where $H$ denotes the Shannon entropy for the measurement
outcomes, i.e., $H(X)=-\sum_x p\,(x)dx\log(p\,(x)dx)$, $H(\bar
A)=-\tr\bar\rho\log\bar\rho$ is the von Neumann entropy of the
measured quantum state $\bar\rho$ and the conditional entropy
$H(A|X)$ is $\sum_x p\,(x)dxS(\rho_{|x})$. In the case of
continuous variables, this expression is not bounded in the limit
$dx\to 0$. Therefore, we will only take such limit (if necessary)
for the computation of the final mutual (or Holevo) information
quantities,
which stay finite. 

Second, for any state $\rho$, one has \cite{Jaynes}
\begin{equation}
\tilde\Delta H(A)=H(\rho\Vert\tilde{\rho})\ge 0 ,\label{funda}
\end{equation}
where $H(\rho\Vert\tilde{\rho})$ denotes the relative entropy
\begin{equation}\label{relent}
    H(\rho\Vert\tilde{\rho})=\tr(\rho\log\rho)-\tr(\rho\log\tilde\rho).
\end{equation}
Note that since the relative entropy is never negative, the state
of maximal entropy for fixed first and second moments is Gaussian
\cite{Jaynes}. In particular, if Alice and Bob share a state
$\rho_{AB}$, they can bound its entropy from its covariance
matrix, that is, $H(\rho_{AB})\leq H(\tilde{\rho}_{AB})$. Using
similar arguments, it can be seen that the same property is
fulfilled by probability distributions, i.e.,
\begin{equation}
    \tilde\Delta H(X)=H(X\Vert\tilde X)
,\label{classic}
\end{equation}
where
\begin{equation}\label{relenet}
    H(X\Vert\tilde X)=\sum_xp\,(x)dx\log\left(\frac{p\,(x)}
    {\tilde p\,(x)}\right) .
\end{equation}

Third, the relative entropy (\ref{relent}) never increases after
the application of a trace-preserving map (or a stochastic map in
the classical case). That is, for any of those maps, denoted by
${\mathcal T}$, and any two states, $\rho_1$ and $\rho_2$,
\begin{equation}
    H(\rho_1||\rho_2)\geq H({\mathcal T}(\rho_1)||{\mathcal
    T}(\rho_2)) .
\end{equation}
This obviously imply
\begin{equation}\label{contr}
    \tilde\Delta H(A)\geq \tilde\Delta H({\mathcal T}(A)) .
\end{equation}
for any Gaussian trace preserving channel $\mathcal T$, and for
any quantum state or classical random variable $A$.

To prove the optimality of Gaussian attacks, we first show that for fixed first and second moments, the Gaussian
attack maximizes Eve's information, $\chi(X:E)$. In order to give the maximally possible information to Eve, one
has to consider that
the global state shared by Alice, Bob and Eve is pure. Then,
\begin{align}
\label{diffhol}
\tilde\Delta\chi(X:E)
  &= \tilde\Delta H(E) - \tilde\Delta H(E|X) \nonumber\\
  &= \tilde\Delta H(AB) - \tilde\Delta H(AB|X) \nonumber\\
  &= \tilde\Delta H(AB) - \tilde\Delta H(\overline{AB}) +\tilde\Delta
  H(X) ,
\end{align}
where we first use the fact that the global state is pure and then \eqref{descomp}. Now, since the channel $AB
\rightarrow \overline{AB}$ defined by the $X$-measurement is Gaussian, $\tilde\Delta H(AB) - \tilde\Delta
H(\overline{AB})$ is not negative. This, together with (\ref{funda}), implies that
\begin{equation}
    \tilde\Delta\chi(X:E)=\tilde\chi(X:E)-\chi(X:E)\geq 0 ,
\end{equation}
so the Gaussian attack maximizes Eve's information for fixed first
and second moments.

Furthemore, the mutual information between Alice and Bob is minimized if Eve's
attack is Gaussian: one has
\begin{equation}
\tilde\Delta I(X:Y)=\tilde\Delta H(X) +\tilde\Delta H(Y) -\tilde\Delta H(XY)\le0
\end{equation}
The first term is null since Alice's modulation is Gaussian, and
 the difference of the last two terms is
negative, following from (\ref{contr}), for the map $XY\rightarrow Y$.
The optimality of Gaussian attacks is therefore proved.
A very similar
argument can be used to prove the optimality of these attacks with
respect to Eq. (\ref{Devetak}).

It is important to stress here that most of the known bounds on
the secret-key rate, including Eqs. \eqref{Renner} and
\eqref{Devetak}, were introduced for finite-dimensional systems,
so in principle they should be carefully applied to the continuous
case. However, in Ref. \cite{vAIC}, it is shown that the
sliced-reconciliation CV protocol of \cite{sliced} achieves the
rate \eqref{Renner} for the case of collective attacks. This
result has to be combined with the fact that, for discrete
variable systems as well as for continuous variable systems,
collective attacks are the most powerful general attacks,
\cite{PhDRenner}. This means that the bounds considered in this
article actually provide general security bounds for CV systems.
The explicit computation of these bounds for the Gaussian case,
now proven to be optimal, can be found in \cite{us}.

Before concluding, we would like to comment on the recent related
results of \cite{WGC}. There, it was shown that, for a given
covariance matrix, the state with minimal distillable secret-key
rate is Gaussian, assuming the distillable secret-key rate is a
continuous functional. This implies that, up to the continuity
assumption, Alice and Bob, for fixed first and second moments, can
safely assume their state to be Gaussian, whenever they are able
to apply any protocol. This result is very interesting and
satisfactory from a theoretical point of view. However, one should
be careful when applying it to a practical scenario. Indeed, the
distillable secret-key rate is defined with respect to the optimal
protocol. However, the optimal protocol can be very challenging
from a practical point of view. For instance, it may include local
coherent and non Gaussian operations among several copies of the
state. In particular, it may be quite different from the realistic
protocol considered here, where the techniques (measurements) used
for the correlation distribution are fixed, and experimentally
feasible. 
Thus, one cannot directly apply the results of \cite{WGC} to the
considered protocols and conclude that the optimal collective
attack is Gaussian.

We have studied the security limits for the CV QKD protocols
proposed in \cite{coher} and \cite{het}, using the recently
obtained lower bounds on the secret-key rate under collective and
general attacks, and we have proven the optimality of Gaussian
attacks for these bounds.

In order to improve the derived security conditions, note that we
have always studied the situation in which Alice and Bob use
one-way reconciliation protocols. Two-way communication protocols
should be analyzed as well, to completely solve the problem of
secret key extraction. Such protocols (\emph{e.g.} CASCADE
\cite{CASCADE}) have already being used in key distribution
experiments \cite{exper} or in the scheme proposed in
\cite{coher2}, even if the security analysis for these cases is
only preliminary yet.

{\sl Note added:} The optimality of Gaussian attacks has been also
proven using different techniques in~\cite{Cerf}.

FG's research was supported by a Marie Curie Intra European
Fellowship within the 6th European Community Framework Programme
(Contract No. MEIF-CT-2003-502045). MN and AA acknowledge
financial support from the Spanish MEC, under a ``Ram\'on y Cajal"
grant, and the Fundaci\'on Ram\'on Areces.

\end{document}